\documentclass[10pt,conference]{IEEEtran}
\IEEEoverridecommandlockouts

\usepackage{amsmath,amssymb,amsfonts}
\usepackage{algorithmic}
\usepackage{textcomp}
\usepackage{xcolor}
\usepackage{float}
\usepackage{algorithm}
\usepackage{booktabs}
\usepackage{subfig}
\usepackage{adjustbox}  

\def\BibTeX{{\rm B\kern-.05em{\sc i\kern-.025em b}\kern-.08em
    T\kern-.1667em\lower.7ex\hbox{E}\kern-.125emX}}
\begin{document}

\title{\textbf{Leveraging Kernel Symmetry for Joint Compression and Error Mitigation in Edge Model Transfer}}

\author{
\thanks{Accepted for presentation at IEEE WCNC 2026.}
\IEEEauthorblockN{1\textsuperscript{st} Anis Hamadouche}
\IEEEauthorblockA{\textit{School of Engineering \& Physical Sciences} \\
\textit{Heriot-Watt University}\\
Edinburgh EH14 4AS, UK \\
anis.hamadouche@hw.ac.uk}
\and
\IEEEauthorblockN{2\textsuperscript{nd} Mathini Sellathurai}
\IEEEauthorblockA{\textit{School of Engineering \& Physical Sciences} \\
\textit{Heriot-Watt University}\\
Edinburgh EH14 4AS, UK \\
m.sellathurai@hw.ac.uk}
}

\maketitle


\begin{abstract}
This paper investigates communication-efficient neural network transmission by exploiting structured symmetry constraints in convolutional kernels. Instead of transmitting all model parameters, we propose a degrees-of-freedom (DoF) based codec that sends only the unique coefficients implied by a chosen symmetry group, enabling deterministic reconstruction of the full weight tensor at the receiver. The proposed framework is evaluated under quantization and noisy channel conditions across multiple symmetry patterns, signal-to-noise ratios, and bit-widths. To improve robustness against transmission impairments, a projection step is further applied at the receiver to enforce consistency with the symmetry-invariant subspace, effectively denoising corrupted parameters. Experimental results on MNIST and CIFAR-10 using a DeepCNN architecture demonstrate that DoF-based transmission achieves substantial bandwidth reduction while preserving significantly higher accuracy than pruning-based baselines, which often suffer catastrophic degradation. Among the tested symmetries, \textit{central-skew symmetry} consistently provides the best accuracy-compression tradeoff, confirming that structured redundancy can be leveraged for reliable and efficient neural model delivery over constrained links.
\end{abstract}


\section{Introduction}
Distributing learned neural networks across bandwidth- and energy-limited wireless links is becoming routine in edge and federated settings, where data remain on devices and models (or their updates) must be communicated reliably to many nodes. In federated learning (FL), for example, communication is repeatedly identified as the principal bottleneck: even modest CNNs incur large payloads when exchanged round-by-round, and noisy channels further degrade the delivered parameters and downstream accuracy. These pressures are amplified in IoT deployments and over-the-air (OTA) aggregation schemes that exploit the wireless medium for simultaneous model update transmission, where distortion and packetization effects couple directly to model quality. Consequently, there is a growing need for architectures and protocols that reduce the number of transmitted degrees of freedom (DoF) and simultaneously mitigate channel-induced corruption during model dissemination \cite{mcmahan2017communication}.

We address the challenge of joint model compression and communication reliability by exploiting structural symmetries in convolutional kernels. Prior work on group-equivariant and steerable CNNs has shown that tying parameters across rotations and reflections preserves inductive biases with negligible overhead \cite{cohen2016group,weiler2019general}. In this work, we convert these representational symmetries into a communication advantage. By transmitting only orbit representatives (degrees of freedom, DoF) and reconstructing full kernels at the receiver, the bit payload is reduced in direct proportion to the orbit count. A final symmetry projection enforces consistency across tied entries and simultaneously averages away independent channel perturbations, thereby acting as an implicit, architecture-level error correction.  

The contributions of this paper are as follows. First, we propose symmetry projection as a dual-purpose mechanism that simultaneously compresses convolutional weights and suppresses transmission noise. Second, we design a DoF-based codec that supports a broad set of 2-D kernel symmetries, including a dedicated antisymmetric construction for central-skew constraints, ensuring exact reconstruction prior to projection. Third, we provide extensive empirical validation across multiple symmetries, quantization levels, and channel qualities using a packetized wireless model with BER-driven bit corruption, CRC32 packet integrity checks, and drop-and-zero-fill reconstruction. Our results show that, at moderate and high SNR, DoF transmission achieves baseline accuracy with up to two-thirds fewer bits, while full-parameter transmission also benefits from projection-induced denoising. Fourth, we present theoretical analysis that quantifies compression as a function of orbit count, bounds the computational overhead of projection relative to convolution, and establishes orbit-size–dependent variance reduction. Together, these results situate symmetry-projected DoF Codec as a complementary primitive to pruning and quantization, targeting the emerging intersection of efficient deep learning and reliable edge communication.

The remainder of the paper is organized as follows. Section~II reviews related work on symmetry-aware CNNs, wireless/OTA learning, and model compression. Section~III introduces the wireless channel model used for model transmission. Section~IV presents the proposed symmetry projectors and the neural codecs with receive-side projection. Section~V details the experimental setup. Section~VI reports results and ablations across symmetries and SNRs. Section~VII concludes and outlines directions for future work.

\section{Related Work}
The use of symmetry as an inductive bias in convolutional networks is well established. Group-equivariant CNNs (G-CNNs)~\cite{cohen2016group} generalize convolution to discrete symmetry groups (e.g., rotations and reflections), yielding layers whose feature maps transform predictably under group actions and therefore enjoy increased weight sharing without parameter growth. Foundational results show that such layers can achieve state-of-the-art performance on rotated variants of image benchmarks with negligible overhead for finite groups, establishing group averaging and orbit-tying as principled mechanisms for parameter efficiency. Steerable CNNs~\cite{cohen2016steerable} extend this perspective with a representation-theoretic construction that parameterizes filters within steerable bases, controlling the cost of a filter bank by feature “types.” Harmonic Networks~\cite{worrall2017harmonic} realize global rotation equivariance by constraining filters to circular harmonics, further demonstrating that structured filters can improve accuracy while reducing effective degrees of freedom.

Beyond strict equivariance, several works enforce kernel-level symmetry directly. SymNet~\cite{dzhezyan2021symmetrical} studies explicit axial and point-reflection constraints within convolutional filters and reports competitive accuracy with fewer parameters by tying weights across symmetric stencil positions. More recently, Symmetry-Structured CNNs~\cite{maduranga2023symmetry} derive parameterizations and update rules that maintain symmetric feature maps throughout training, reporting gains (with fewer parameters) on tasks where pairwise interactions induce inherent spatial symmetry (e.g., RNA/protein contact maps, recommendation). These efforts inform our choice of symmetry projectors and underscore the empirical value of tying within-kernel orbits.

Structured linear operators~\cite{sindhwani2015structured} offer a complementary route to compact models. Low-displacement-rank (LDR) matrices—including Toeplitz-like and circulant transforms—compress fully connected or convolution-like mappings while admitting fast multiply and gradient evaluation. Such transforms trade a small bias for substantial parameter and compute savings, and they motivate our inclusion of Toeplitz symmetry as a kernel prior that reduces the number of independent stencil coefficients.

In parallel, communication-efficient learning targets the bandwidth bottleneck in distributed and federated settings. FedAvg~\cite{collins2022fedavg} reduces rounds by aggregating model updates on devices but still faces large payloads for modern networks. Gradient-sparsification and compression methods (e.g., Deep Gradient Compression)~\cite{lin2017deep} shrink update traffic by 100×–600× while preserving convergence through momentum correction and related techniques. Over-the-air (analog) aggregation~\cite{yang2020federated} exploits the superposition property of the wireless channel to sum updates in the air, improving spectral efficiency but exposing models to channel noise and distortion. Our work diverges from these strands by addressing model dissemination itself (rather than gradient exchange) and by using architectural symmetry as an intrinsic compressor and as an implicit error corrector at the receiver via projection, which averages away independent channel perturbations within tied orbits.

Finally, approximate-equivariance formulations~\cite{wang2022approximately} study settings where symmetries hold only approximately, showing that relaxed constraints can retain the benefits of inductive bias under real-world imperfections. This perspective aligns with our empirical observation that receive-side projection onto a symmetry subspace often improves accuracy by denoising channel-perturbed weights, even when the transmitted kernels have deviated from perfect symmetry.

\section{Channel Model}
We model the weight-transmission link as a memoryless binary-input additive white Gaussian noise (AWGN) channel with BPSK signaling. For a target signal-to-noise ratio (SNR) in dB, the corresponding bit error rate (BER) is
\begin{equation}
\label{eq:ber_bpsk}
\mathrm{BER}(\mathrm{SNR}_{\mathrm{dB}}) \;=\; Q\!\Big(\sqrt{2 \cdot 10^{\mathrm{SNR}_{\mathrm{dB}}/10}}\Big),
\end{equation}
where $Q(\cdot)$ is the standard Gaussian tail function. Bit errors are injected independently per bit according to~\eqref{eq:ber_bpsk} (no burst model).

Payloads are packetized into fixed-length frames of $2048$ bits. Each packet contains a big-endian header and a CRC footer:
\begin{itemize}
\item \emph{Header (10 bytes)}: \textit{[u32 stream\_id][u16 layer\_id][u32 seq\_idx]}.
\item \emph{Body}: contiguous slice of the serialized payload.
\item \emph{Footer (4 bytes)}: CRC32 over the header+body.
\end{itemize}
Packets are independently corrupted at the bit level by the BER process; optional packet erasures are disabled ($p_{\mathrm{loss}}{=}0$). A packet is accepted only if its CRC32 matches; failed-CRC packets are discarded. No ARQ/interleaving or FEC is used. At the receiver, packets are reassembled in sequence number order; missing segments (due to loss or CRC failure) are zero-filled. The effective per-packet error rate (PER) and delivered fraction are logged.

Each convolutional layer is serialized as follows. Given a symmetry type $S$, we extract orbit representatives (degrees of freedom, DoF) from the FP32 kernel, uniformly quantize them to signed $b$-bit integers ($b\in\{8,16\}$) with symmetric per-tensor scaling, and serialize to bytes. On reception, bytes are dequantized and used to reconstruct the full kernel by populating each orbit with its representative; for central-skew symmetry, a special antisymmetric pairing enforces a zero center coefficient. A final \emph{symmetry projection} (same operator used during training) can be applied at runtime to enforce exact constraints and average independent perturbations across each orbit.

We emphasize that no retransmission mechanism (ARQ), interleaving, or forward error correction (FEC) is applied in the reported experiments; the receiver performs a single-pass decode where packets failing CRC are discarded and their payload segments are replaced by zeros.

\section{Proposed Symmetry–Projection and DoF Codec}
\label{sec:method}

We formalize two complementary mechanisms: (i) a \emph{symmetry projection} applied during training and inference, and (ii) a \emph{degrees–of–freedom (DoF) codec} used during transmission. Together they exploit weight tying induced by 2-D kernel symmetries to reduce payload size and to suppress channel/quantization noise via receive-side projection.

\subsection{Symmetries, Orbits, and Projection}
Let $W\!\in\!\mathbb{R}^{C_{\text{out}}\times C_{\text{in}}\times K\times K}$ denote a convolutional stencil with odd $K$. A symmetry $S$ acts on spatial indices $(i,j)\!\in\!\{0,\ldots,K\!-\!1\}^2$ through rotations, reflections, or transposition, partitioning the index set into disjoint \emph{orbits} $\{\mathcal{O}_m\}_{m=1}^{M(S)}$. Entries within an orbit are constrained to be equal (even) or equal up to sign (skew).

If $S$ is induced by a finite group $G$ acting on stencils, the (Reynolds) projector is
\begin{equation}
\mathcal{P}_{G}(W)\;=\;\frac{1}{|G|}\sum_{g\in G} g(W), 
\qquad \mathcal{P}_{G}^{2}=\mathcal{P}_{G}.
\end{equation}
For non-group partitions (e.g., radial bins or Toeplitz diagonals), projection averages within each orbit,
\begin{equation}
\big[\mathcal{P}_{S}(W)\big]_{i,j}
\;=\;\frac{1}{|\mathcal{O}_m|}\!\sum_{(u,v)\in\mathcal{O}_m}\! W_{u,v},
\qquad (i,j)\in\mathcal{O}_m .
\end{equation}

\begin{figure}[!t]
  \centering
  \adjustbox{rotate=90,max width=\columnwidth}{%
    \begin{minipage}{\linewidth}
      \centering
      \footnotesize
      \subfloat[Central-even]{\includegraphics[width=0.23\linewidth]{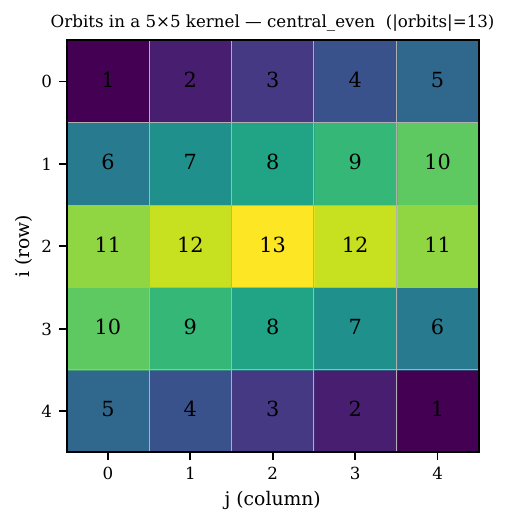}\label{fig:orbit-central-even}}\hfill
      \subfloat[Central-skew]{\includegraphics[width=0.23\linewidth]{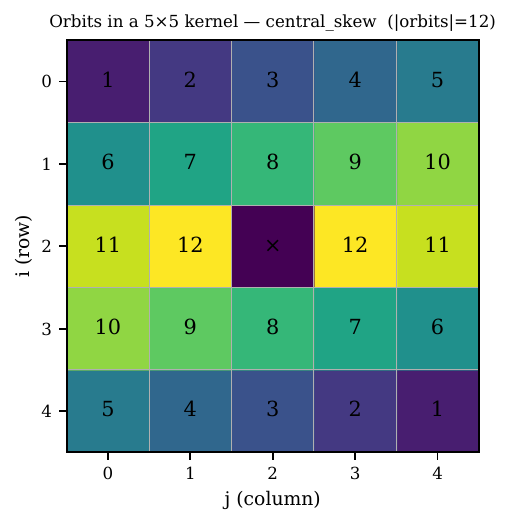}\label{fig:orbit-central-skew}}\hfill
      \subfloat[Horizontal mirror]{\includegraphics[width=0.23\linewidth]{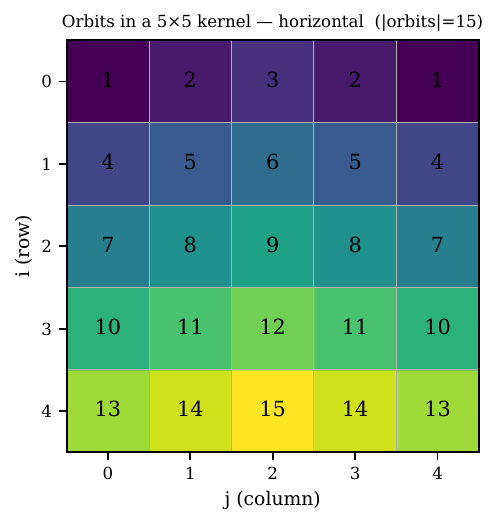}\label{fig:orbit-horizontal}}\hfill
      \subfloat[Main diagonal]{\includegraphics[width=0.23\linewidth]{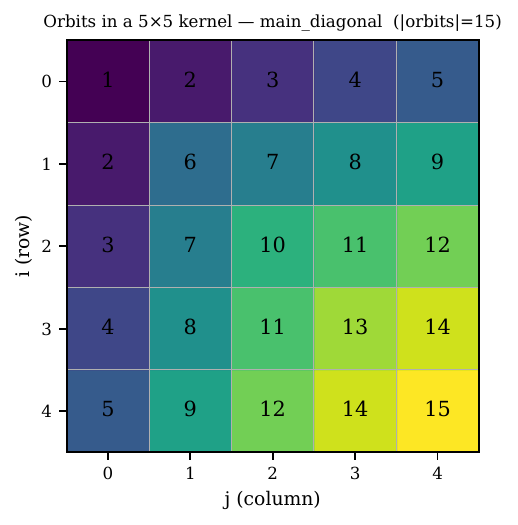}\label{fig:orbit-maindiag}}\\[2pt]
      \subfloat[Quarter rotations (C4)]{\includegraphics[width=0.23\linewidth]{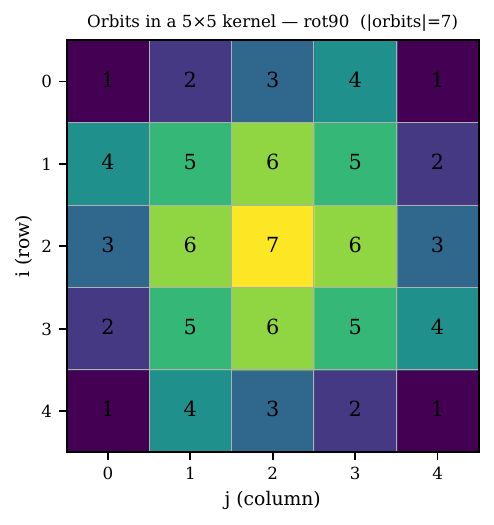}\label{fig:orbit-rot90}}\hfill
      \subfloat[Radial invariance]{\includegraphics[width=0.23\linewidth]{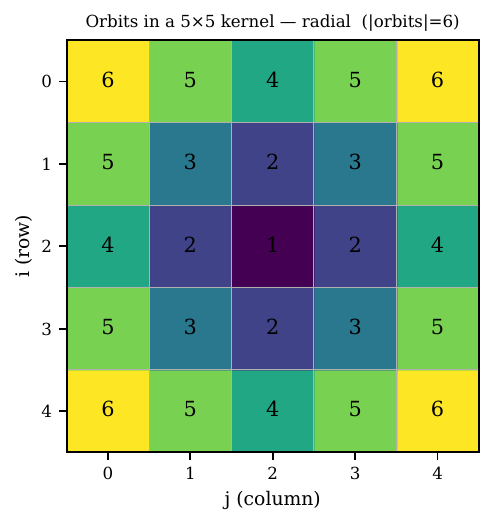}\label{fig:orbit-radial}}\hfill
      \subfloat[Toeplitz (constant diagonals)]{\includegraphics[width=0.23\linewidth]{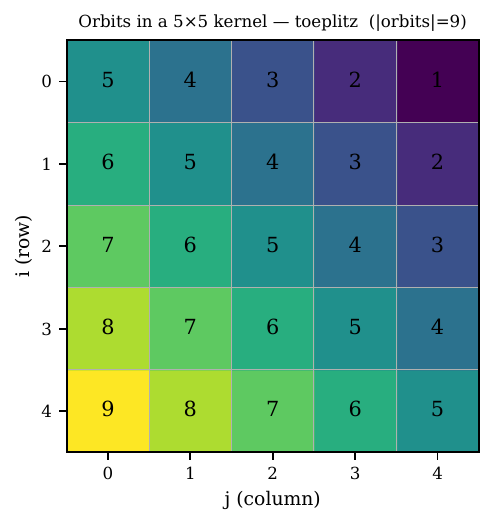}\label{fig:orbit-toeplitz}}
    \end{minipage}%
  }
  \caption{Orbits in a $5\times 5$ convolution kernel under different symmetry groups. Each panel displays the orbit ID assigned to index $(i,j)$ (darker/lighter cells denote different IDs). Entries with the same ID belong to the same orbit and are tied to a single degree of freedom. In \protect\subref{fig:orbit-central-skew}, the center element is constrained by antisymmetry and marked with $\times$.}
 \label{fig:orbit-gallery}
\end{figure}

Example orbit decompositions for common symmetries are illustrated in Fig.~\ref{fig:orbit-gallery}.

\paragraph*{Noise–reduction intuition}
Suppose independent zero-mean perturbations within orbit $\mathcal{O}_m$ with variance $\sigma^2$. Averaging $n_m\!=\!|\mathcal{O}_m|$ samples reduces the variance to $\sigma^2/n_m$, hence
\begin{equation}
\mathbb{E}\!\left[\|\widehat{W}-W\|_{F}^{2}\right]
\;\approx\; C_{\text{out}}C_{\text{in}}\sum_{m=1}^{M(S)}\frac{\sigma^{2}}{n_m}
\;\le\; C_{\text{out}}C_{\text{in}}\,M(S)\,\sigma^{2},
\end{equation}
so larger orbits yield stronger denoising. This benefit applies to both full model transmission and DoF reconstructions.

\subsection{Use in the Forward Pass}
Each \texttt{SymConv2d} layer evaluates
\begin{equation}
Y\;=\;\mathrm{Conv}\!\big(\mathcal{P}_{S}(W),\,X\big),
\end{equation}
thereby enforcing the constraint at train and test time. The projection overhead scales as $\Theta\!\big(C_{\text{out}}C_{\text{in}}K^{2}c_{S}\big)$, where $c_{S}$ is the number of transforms averaged (a small constant). Which is negligible compared with the convolution cost. 

During both training and inference, the convolution operates on the projected weights $\mathcal{P}_S(W)$ rather than on $W$ directly:
\begin{equation}
Y = \mathrm{Conv}\!\left(\mathcal{P}_S(W), X\right),
\end{equation}
where $X$ is the input activation and $Y$ is the output.

Backpropagation through the projection is straightforward because $\mathcal{P}_S$ is linear. For any loss function $\mathcal{L}$, the gradient with respect to the unprojected kernel $W$ is obtained by
\begin{equation}
\nabla_W \mathcal{L} \;=\; \mathcal{P}_S^\top \nabla_{\mathcal{P}_S(W)} \mathcal{L},
\end{equation}
where $\mathcal{P}_S^\top$ is the adjoint operator of $\mathcal{P}_S$.  
Since $\mathcal{P}_S$ is an orthogonal projector, $\mathcal{P}_S^\top = \mathcal{P}_S$, implying that gradients are tied across all entries within the same orbit. This ensures that the update direction respects the imposed symmetry.

The overall training loop follows standard stochastic optimization with weight projection at every forward pass. Algorithm~\ref{alg:sym-training} summarizes the process.  

\begin{algorithm}[!b]
\caption{Training CNN with Symmetry Projection}
\label{alg:sym-training}
\begin{algorithmic}[1]
\STATE Initialize network parameters $\{W_\ell\}$ for each layer $\ell$.
\FOR{epoch $=1,\dots,E$}
  \FOR{each minibatch $(X,Y)$}
    \FOR{each convolutional layer $\ell$}
      \STATE Project weights: $\widetilde{W}_\ell = \mathcal{P}_{S_\ell}(W_\ell)$
      \STATE Forward propagate using $\widetilde{W}_\ell$
    \ENDFOR
    \STATE Compute loss $\mathcal{L}$
    \STATE Backpropagate gradients through $\mathcal{P}_{S_\ell}$
    \STATE Update $\{W_\ell\}$ with optimizer (e.g., AdamW)
  \ENDFOR
\ENDFOR
\end{algorithmic}
\end{algorithm}


\subsection{DoF Encoder (Bandwidth Compression)}
For transmission, we serialize a single representative per orbit. Let $s\!\in\!\mathbb{R}^{C_{\text{out}}C_{\text{in}}M(S)}$ collect orbit averages (or signed representatives for skew):
\begin{equation}
s_m \;=\; \frac{1}{|\mathcal{O}_m|}\sum_{(u,v)\in\mathcal{O}_m} \big(T(W)\big)_{u,v},
\qquad m=1,\dots,M(S),
\end{equation}
where $T(\cdot)$ denotes (de)quantization/serialization. With $b$-bit signed quantization, the \emph{full} and \emph{DoF} layer payloads are
\begin{equation}
B_{\text{full}} \;=\; b\,C_{\text{out}}C_{\text{in}}K^{2},\qquad
B_{\text{dof}} \;=\; b\,C_{\text{out}}C_{\text{in}}M(S),
\end{equation}
yielding a bandwidth saving
\begin{equation}
\eta(S)\;=\;1-\frac{B_{\text{dof}}}{B_{\text{full}}}
\;=\;1-\frac{M(S)}{K^{2}}.
\end{equation}

\subsection{DoF Decoder and Receive–Side Projection}
At the receiver, a stencil is synthesized from $s$ and then projected:
\begin{equation}
\begin{split}
\big[\mathrm{Synth}_{S}(s)\big]_{i,j}&=
\begin{cases}
s_m, & (i,j)\in\mathcal{O}_m \ \text{(even)},\\
\pm s_m, & (i,j)\in\mathcal{O}_m \ \text{(skew)},
\end{cases}
\\
\widehat{W}\;&=\;\mathcal{P}_{S}\!\big(\mathrm{Synth}_{S}(s)\big).
\end{split}
\end{equation}
Since $\mathcal{P}_{S}$ is the orthogonal projector onto the invariant subspace, it is mean-square optimal and further suppresses channel and quantization noise.

\subsection{Central–Skew Codec (Signed Pairs)}
For $K=2c+1$, define the skew subspace
\[
S_{\text{skew}}=\Big\{W:\ W_{i,j}=-W_{K-1-i,\,K-1-j},\ \ W_{c,c}=0\Big\}.
\]
There are $M=\tfrac{K^{2}-1}{2}$ independent $180^\circ$ pairs. The linear, bijective map
\begin{equation}
\begin{split}
&\Phi:\ \mathbb{R}^{M}\!\to S_{\text{skew}},\\
&\Phi(s)\Big|_{\{(i,j),(K-1-i,K-1-j)\}}=(+s_m,-s_m)\\ 
&\Phi(s)_{c,c}=0,
\end{split}
\end{equation}
satisfies
\begin{equation}
\|\Phi(s)\|_{F}^{2}=2\|s\|_{2}^{2},
\qquad
\|\Phi(s)-\Phi(t)\|_{F}=\sqrt{2}\,\|s-t\|_{2},
\end{equation}
so $\widetilde{\Phi}=\Phi/\sqrt{2}$ is an isometry. This conditioning improves robustness to quantization and bit errors. Moreover, $\mathcal{P}_{\text{skew}}(\Phi(s))=\Phi(s)$ (idempotence), implying zero post-projection distortion in the absence of channel/quantization noise. 

Projection becomes beneficial only when additional symmetry-breaking perturbations occur after reconstruction (e.g., receiver noise, quantization mismatch, or storage corruption), in which case it acts as a denoising step that restores exact symmetry.

\subsection{Reliability vs. Bandwidth}
With fixed packet length $L$ bits and per-packet success probability $p_{\text{succ}}\!\in\!(0,1)$, the probability that all $N\!=\!\lceil B/L\rceil$ packets for a layer are delivered is
\begin{equation}
\Pr\{\text{layer clean}\}=p_{\text{succ}}^{\,N}.
\end{equation}
Because $B_{\text{dof}}<B_{\text{full}}$ whenever $M(S)<K^{2}$, DoF transmission uses fewer packets, strictly increasing $\Pr\{\text{layer clean}\}$ at fixed channel quality. Coupled with receive-side projection, this explains the observed gains of symmetry-preserving neural codecs, with DoF additionally enjoying reduced payload and enhanced delivery reliability.

\section{Experimental Setup}
\label{sec:exp_setup}

We evaluate the proposed DoF-based wireless transmission framework on two standard image classification benchmarks: MNIST and CIFAR-10. For MNIST, models are trained for $8$ epochs using a batch size of $128$, while for CIFAR-10 models are trained for $30$ epochs using the same batch size. For CIFAR-10, standard data augmentation is applied during training (random cropping with padding and random horizontal flipping), followed by per-channel normalization. All reported results are averaged over $10$ independent runs.

We consider a DeepCNN architecture consisting of three convolutional layers followed by a global pooling classification head. Each convolutional kernel is constrained to satisfy a selected symmetry by applying a projection operator $\mathcal{P}_{\mathrm{sym}}(\cdot)$ during the forward pass. We evaluate the following symmetry families:
\textit{none}, \textit{central-even}, \textit{central-skew}, \textit{horizontal}, \textit{vertical}, \textit{main-diagonal}, \textit{anti-diagonal}, \textit{rot90}, \textit{radial}, and \textit{toeplitz}. 
For each symmetry, the model is trained independently from scratch under the corresponding constraint.

For a given symmetry $\mathrm{sym}$, each convolutional kernel $\mathbf{W}$ is represented by a reduced set of unique degrees of freedom (DoF) coefficients, denoted by $\mathbf{d}_{\mathrm{sym}} \in \mathbb{R}^{U_{\mathrm{sym}}}$, where $U_{\mathrm{sym}}$ is determined by the orbit structure induced by the symmetry group. The DoF coefficients are uniformly quantized using symmetric $b$-bit quantization, with $b = 8$ for MNIST and CIFAR-10. The receiver reconstructs the full kernel by expanding the quantized DoF coefficients and applies receive-side projection to enforce the symmetry manifold.

Wireless transmission is simulated over an AWGN channel with BPSK modulation. The bit error rate (BER) is derived from the signal-to-noise ratio (SNR) using the standard $Q$-function model. We evaluate at $\mathrm{SNR}=10$~dB. The DoF payload is packetized into fixed-length packets, and each packet is protected with a CRC checksum. Packets failing CRC are dropped and replaced by zeros at the receiver. Automatic repeat request (ARQ) retransmissions are disabled by default, such that the evaluation reflects a single-shot unreliable wireless delivery.

To provide a fair sparsity-based baseline, we additionally construct a matched pruning equivalent for each symmetry setting by pruning the unconstrained (\textit{none}) model to match the same DoF transmission budget. Let $\mathbf{w}\in\mathbb{R}^{N}$ denote the vectorized convolutional parameters of the \textit{none} model, and let $U_{\mathrm{sym}}$ denote the number of transmitted DoF coefficients under a given symmetry constraint. We define the pruning budget as
\begin{equation}
k = U_{\mathrm{sym}},
\end{equation}
and perform fixed-NNZ magnitude pruning by retaining the indices corresponding to the $k$ largest-magnitude entries:
\begin{equation}
\mathcal{I}_k = \arg\max_{\substack{\mathcal{I}\subset\{1,\dots,N\} \\ |\mathcal{I}|=k}}
\sum_{i\in\mathcal{I}} |w_i|.
\end{equation}
The resulting sparse pruned weight vector $\tilde{\mathbf{w}}\in\mathbb{R}^{N}$ is then defined as
\begin{equation}
\tilde{w}_i =
\begin{cases}
w_i, & i \in \mathcal{I}_k, \\
0, & \text{otherwise}.
\end{cases}
\end{equation}
For transmission, the sparse tensor is serialized in COO format by transmitting index-value pairs $\{(i,\tilde{w}_i)\}_{i\in\mathcal{I}_k}$, where values are quantized to the same bit-width $b$. This explicitly accounts for sparse index overhead and provides a direct comparison between structured DoF compression and sparsity-based compression under identical wireless conditions.

We report classification accuracy at the transmitter (Acc$_{\mathrm{Tx}}$), quantized pre-channel accuracy (Acc$_{\mathrm{Tx,q}}$), and post-transmission receiver accuracy after reconstruction and projection (Acc$_{\mathrm{Rx}}$). Performance is summarized using $\Delta$Acc, measured in percentage points relative to the post-transmission baseline (\textit{none}) at the same bit-width. We additionally report the number of transmitted unique parameters (DoF budget), the resulting payload size, payload reduction relative to the full baseline transmission, the total number of bits sent over the packetized channel (including CRC and retransmission overhead if enabled), and the end-to-end (E2E) latency from transmitter payload formation to receiver reconstruction and projection. All results are reported as mean(SE) over $10$ runs.

\section{Results and Discussion}

\begin{table*}[t]
\vspace{0.08in}
\centering
\resizebox{\textwidth}{!}{%
\begin{tabular}{|l c c c c c c c c c|}
\hline
Sym / Pruned &
Acc$_{\mathrm{Tx}}$ (\%) &
Acc$_{\mathrm{Tx,q}}$ (\%) &
Acc$_{\mathrm{Rx}}$ (\%) &
$\Delta$Acc (pp) &
Unique ($\times 10^{3}$) &
Payload (kbits) &
Payload Reduct. (\%) &
Bits Sent (kbits) &
E2E Latency (ms) \\
\hline
none / pruned eq.
& 98.2 / 98.2
& 98.3 / 98.2
& 93.2(4.17) / 97.8(0.18)
& 0.00 / +4.54
& 92.4 / 92.4
& 739.6 / 3698.0
& 0.00 / -400.01
& 786.4 / 3913.7
& 75.8(6.01) / 584.1(14.3) \\

central-even / pruned eq.
& 90.8 / 98.2
& 89.6 / 98.2
& 87.1(1.36) / 56.5(1.61)
& -6.17 / -36.74
& 51.4 / 51.4
& 410.9 / 2054.5
& 44.44 / -177.79
& 438.3 / 2177.0
& 39.6(2.41) / 330.1(9.70) \\

central-skew / pruned eq.
& \textbf{98.8} / 98.2
& \textbf{98.8} / 98.2
& \textbf{98.8(0.02)} / 32.1(0.90)
& \textbf{+5.53} / -61.11
& 41.1 / 41.1
& \textbf{328.7} / 1643.6
& \textbf{55.56} / -122.24
& \textbf{350.2} / 1742.8
& \textbf{32.5(1.73)} / 256.9(6.33) \\

horizontal / pruned eq.
& 95.6 / 98.2
& 95.1 / 98.2
& 91.0(3.18) / 89.1(0.41)
& -2.24 / -4.16
& 61.6 / 61.6
& 493.1 / 2465.4
& 33.33 / -233.35
& 524.3 / 2609.2
& 41.2(0.99) / 371.1(7.27) \\

vertical / pruned eq.
& 94.4 / 98.2
& 93.8 / 98.2
& 82.2(7.52) / 89.1(0.41)
& -11.04 / -4.16
& 61.6 / 61.6
& 493.1 / 2465.4
& 33.33 / -233.35
& 524.3 / 2609.2
& 44.3(1.96) / 358.0(5.59) \\

main-diagonal / pruned eq.
& 95.4 / 98.2
& 95.7 / 98.2
& 89.9(3.74) / 89.1(0.41)
& -3.32 / -4.16
& 61.6 / 61.6
& 493.1 / 2465.4
& 33.33 / -233.35
& 524.3 / 2609.2
& 40.1(0.73) / 386.4(11.2) \\

anti-diagonal / pruned eq.
& 93.2 / 98.2
& 93.3 / 98.2
& 92.6(0.35) / 89.1(0.41)
& -0.63 / -4.16
& 61.6 / 61.6
& 493.1 / 2465.4
& 33.33 / -233.35
& 524.3 / 2609.2
& 41.6(1.13) / 390.1(5.64) \\

rot90 / pruned eq.
& 77.9 / 98.2
& 78.2 / 98.2
& 70.1(5.41) / 9.8(0.00)
& -23.08 / -83.42
& \textbf{30.8} / \textbf{30.8}
& \textbf{246.5} / 1232.7
& \textbf{66.67} / -66.68
& \textbf{264.2} / 1304.6
& 27.3(1.34) / 219.8(4.53) \\

radial / pruned eq.
& 72.6 / 98.2
& 73.5 / 98.2
& 67.7(4.38) / 9.8(0.00)
& -25.52 / -83.42
& \textbf{30.8} / \textbf{30.8}
& \textbf{246.5} / 1232.7
& \textbf{66.67} / -66.68
& \textbf{264.2} / 1304.6
& 26.9(1.87) / 206.1(4.85) \\

toeplitz / pruned eq.
& 95.2 / 98.2
& 95.3 / 98.2
& 85.9(5.98) / 56.5(1.61)
& -7.36 / -36.74
& 51.4 / 51.4
& 410.9 / 2054.5
& 44.44 / -177.79
& 438.3 / 2177.0
& 40.8(2.91) / 298.5(3.63) \\

\hline
\end{tabular}}
\caption{MNIST DeepCNN wireless evaluation at $\mathrm{SNR}=10$~dB for $b=8$ bits (mean(SE) over 10 runs). Each entry reports the symmetry result followed by its pruned equivalent (value / pruned-value). Acc$_{\mathrm{Tx}}$ is the trained accuracy before transmission, Acc$_{\mathrm{Tx,q}}$ is the DoF-quantized accuracy before transmission, and Acc$_{\mathrm{Rx}}$ is the post-transmission accuracy after reconstruction and receive-side projection. $\Delta$Acc is measured in percentage points (pp) relative to the post-transmission baseline (none). Payload reduction is computed relative to the full baseline payload. Payload and Bits Sent are reported in kbits ($10^{3}$ bits), and Unique is reported in units of $10^{3}$ parameters. Bold values indicate the best performance in each column. E2E Latency (ms) denotes the end-to-end transmission time from DoF/sparse payload formation at the transmitter (Tx), packetized wireless delivery, to full weight reconstruction and symmetry projection at the receiver (Rx).}
\label{tab:mnist_deep_snr10_8bit_sym_pruned}
\end{table*}


\begin{table*}[t]
\vspace{0.08in}
\centering
\resizebox{\textwidth}{!}{%
\begin{tabular}{|l c c c c c c c c c|}
\hline
Sym / Pruned &
Acc$_{\mathrm{Tx}}$ (\%) &
Acc$_{\mathrm{Tx,q}}$ (\%) &
Acc$_{\mathrm{Rx}}$ (\%) &
$\Delta$Acc (pp) &
Unique ($\times 10^{3}$) &
Payload (kbits) &
Payload Reduct. (\%) &
Bits Sent (kbits) &
E2E Latency (ms) \\
\hline
none / pruned eq.
& \textbf{77.7} / 77.7
& \textbf{77.9} / 77.7
& 76.7(0.42) / \textbf{76.8(0.20)}
& 0.00 / \textbf{+0.14}
& 93.0 / 93.0
& 744.2 / 3721.1
& 0.00 / -400.01
& 790.5 / 3938.3
& 61.3(1.71) / 579.5(26.7) \\

central-even / pruned eq.
& 66.1 / 77.7
& 66.1 / 77.7
& 63.1(2.50) / 42.8(0.24)
& -13.55 / -33.85
& 51.7 / 51.7
& 413.4 / 2067.3
& 44.44 / -177.79
& 440.3 / 2189.3
& 34.4(1.82) / 321.0(16.5) \\

central-skew / pruned eq.
& 72.9 / 77.7
& 73.0 / 77.7
& 72.5(0.27) / 23.9(0.07)
& -4.19 / -52.82
& 41.3 / 41.3
& 330.8 / 1653.9
& 55.56 / -122.24
& 352.3 / 1753.1
& 28.2(1.25) / 257.7(13.3) \\

horizontal / pruned eq.
& 72.9 / 77.7
& 72.4 / 77.7
& 71.7(0.26) / 53.0(0.41)
& -5.01 / -23.72
& 62.0 / 62.0
& 496.1 / 2480.7
& 33.33 / -233.35
& 528.4 / 2625.5
& 41.9(1.84) / 398.7(11.9) \\

vertical / pruned eq.
& 64.6 / 77.7
& 64.4 / 77.7
& 63.4(0.56) / 53.0(0.41)
& -13.27 / -23.72
& 62.0 / 62.0
& 496.1 / 2480.7
& 33.33 / -233.35
& 528.4 / 2625.5
& 44.1(2.51) / 371.2(13.2) \\

main-diagonal / pruned eq.
& 67.8 / 77.7
& 68.0 / 77.7
& 67.2(0.25) / 53.0(0.41)
& -9.48 / -23.72
& 62.0 / 62.0
& 496.1 / 2480.7
& 33.33 / -233.35
& 528.4 / 2625.5
& 39.9(1.18) / 395.3(14.5) \\

anti-diagonal / pruned eq.
& 68.8 / 77.7
& 67.8 / 77.7
& 65.8(1.05) / 53.0(0.41)
& -10.86 / -23.72
& 62.0 / 62.0
& 496.1 / 2480.7
& 33.33 / -233.35
& 528.4 / 2625.5
& 42.3(1.36) / 391.1(10.7) \\

rot90 / pruned eq.
& 58.2 / 77.7
& 54.1 / 77.7
& 53.8(0.21) / 18.6(0.26)
& -22.87 / -58.12
& \textbf{31.0} / \textbf{31.0}
& \textbf{248.1} / 1240.4
& \textbf{66.67} / -66.68
& \textbf{266.2} / 1312.8
& 24.3(1.41) / 212.8(5.35) \\

radial / pruned eq.
& 57.4 / 77.7
& 53.5 / 77.7
& 51.8(1.31) / 18.6(0.26)
& -24.86 / -58.12
& \textbf{31.0} / \textbf{31.0}
& \textbf{248.1} / 1240.4
& \textbf{66.67} / -66.68
& \textbf{266.2} / 1312.8
& \textbf{21.0(0.62)} / 219.0(8.24) \\

toeplitz / pruned eq.
& 62.7 / 77.7
& 61.9 / 77.7
& 62.0(0.16) / 42.8(0.24)
& -14.73 / -33.85
& 51.7 / 51.7
& 413.4 / 2067.3
& 44.44 / -177.79
& 440.3 / 2189.3
& 40.3(3.73) / 330.7(14.5) \\
\hline
\end{tabular}}
\caption{CIFAR-10 DeepCNN wireless evaluation at $\mathrm{SNR}=10$~dB for $b=8$ bits (mean(SE) over 10 runs).}
\label{tab:cifar10_deep_snr10_8bit_sym_pruned}
\end{table*}

Across both MNIST (Table~\ref{tab:mnist_deep_snr10_8bit_sym_pruned}) and CIFAR-10 (Table~\ref{tab:cifar10_deep_snr10_8bit_sym_pruned}) at $\mathrm{SNR}=10$~dB and $b=8$ bits, the proposed DoF-based symmetry codec consistently achieves substantial payload reduction while maintaining competitive post-transmission accuracy compared to its pruned equivalent. In particular, \textit{central-skew} provides the best overall accuracy--bandwidth tradeoff on MNIST, achieving the highest post-transmission accuracy (98.8\%) and the largest positive gain over the baseline ($\Delta$Acc $=+5.53$ pp) while reducing the payload by 55.56\% and yielding the lowest end-to-end latency (32.5 ms). On CIFAR-10, although the baseline (\textit{none}) remains the best-performing in terms of absolute accuracy, \textit{central-skew} still achieves strong performance (72.5\% Rx) while providing 55.56\% payload reduction and significantly lower latency (28.2 ms), demonstrating robustness even in a harder dataset.

In contrast, the pruned equivalents consistently exhibit severe overhead due to sparse index transmission, resulting in payload expansion rather than compression (e.g., $-400\%$ for the \textit{none} pruned baseline) and dramatically higher latency (typically $>200$ ms). Moreover, pruning often leads to catastrophic accuracy collapse under wireless corruption, particularly for aggressive compression targets such as \textit{rot90} and \textit{radial}, where pruned equivalents drop to 9.8\% on MNIST and 18.6\% on CIFAR-10. Strong symmetry constraints such as \textit{rot90} and \textit{radial} provide the largest bandwidth savings (66.67\%) and lowest latency, but incur substantial accuracy degradation on both datasets, making them suitable only for extreme bandwidth-limited regimes. Overall, these results confirm that structured DoF transmission via symmetry constraints is significantly more efficient and robust than sparsity-based pruning baselines, particularly when accounting for realistic transmission overhead and end-to-end latency.

\subsection{Practical Guidelines for Symmetry Selection}
In practice, the symmetry constraint should be selected based on the desired accuracy--bandwidth tradeoff and the anticipated severity of wireless channel impairments. If post-transmission accuracy is the primary objective, weak-to-moderate constraints such as \textit{central-skew} are recommended, as they consistently provide the best DoF robustness and the largest positive $\Delta$Acc relative to the baseline while still achieving substantial payload reduction (typically $55.56\%$). If bandwidth minimization is the dominant objective, stronger symmetries such as \textit{rot90} and \textit{radial} yield the highest compression gains (up to $66.67\%$ payload reduction) and the lowest transmission time, but often incur significant accuracy degradation, limiting their applicability to non-critical tasks. Intermediate reflection and diagonal symmetries (e.g., \textit{horizontal}, \textit{vertical}, and diagonal constraints) provide moderate compression ($33.33\%$) but generally offer limited advantage over \textit{central-skew}, especially in more challenging regimes where DoF accuracy becomes less stable.

\section{Conclusion}
This work studied symmetry-aware neural network transmission using a degree-of-freedom (DoF) codec under quantization and noisy wireless channel impairments. By exploiting structured parameter tying induced by symmetry constraints, the proposed framework enables deterministic payload reduction while preserving the functional integrity of convolutional kernels. Experiments on MNIST and CIFAR-10 demonstrate that DoF-based symmetry transmission achieves a favorable accuracy--bandwidth tradeoff and consistently outperforms magnitude-based pruning baselines, which often suffer from severe degradation due to sparse index overhead and instability under channel corruption. Moreover, receive-side symmetry projection acts as an implicit denoiser, correcting symmetry violations introduced during transmission and improving post-reception robustness without requiring additional redundancy. Among the evaluated constraints, \textit{central-skew} emerges as the most effective operating point, offering the strongest balance between compression efficiency and accuracy preservation. Overall, the results highlight structured DoF transmission as a practical alternative to sparsity-driven compression for communication-efficient deployment of deep models in bandwidth-limited and noise-impaired environments.

While the current evaluation considers a memoryless AWGN channel with packet-level CRC dropping, extending the framework to more realistic fading channels, burst-error regimes, and adaptive link-layer retransmission mechanisms remains an important direction for future work.
\section*{Acknowledgment}
This work was supported by the UK Engineering and Physical Sciences Research Council (EPSRC) grant EP/Y037243/1 for TITAN Extension Federated Telecommunications Hub.

\appendices



\bibliographystyle{./IEEEtran}
\bibliography{./myBibIEEE}

\end{document}